\documentclass[reprint,NumberedRefs]{JASAnew}

\usepackage{amsmath}
\usepackage{natbib}	
\usepackage{graphicx,color,rotating}
\usepackage[latin1]{inputenc}
\usepackage{textcomp}
\usepackage{dcolumn}
\usepackage{bm}     
\usepackage{upgreek}
\usepackage{subdepth}  			
\usepackage{multirow}
\usepackage[latin1]{inputenc}

\usepackage{hyperref}						
\usepackage{xcolor}						
\hypersetup{							
    colorlinks,
    linkcolor={blue!80!black},
    citecolor={blue!80!black},
    urlcolor={blue!80!black}
}
\definecolor{darkgreen}{rgb}{0.00, 0.50, 0.00}
\definecolor{DARKGREEN}{rgb}{0.00, 0.50, 0.00} 
\definecolor{RED}{rgb}{1.00, 0.00, 0.00} 
\definecolor{GREEN}{rgb}{0.00, 1.00, 0.00} 
\definecolor{BLUE}{rgb}{0.00, 0.00, 1.00} 
\definecolor{MAGENTA}{rgb}{1.00, 0.00, 1.00} 

\newcommand{\mr}[1]{\ensuremath{\mathrm{#1}}}
\renewcommand{\vec}[1]{\bm{#1}}
\newcommand{\ee}{\mathrm{e}}
\newcommand{\ii}{\mathrm{i}}
\newcommand{\dm}{\mathrm{d}}
\newcommand{\avr}[1]{\big\langle #1 \big\rangle}

\DeclareMathOperator{\re}{Re}

\newcommand{\iot}{{\ii\omega t}}

\newcommand{\pp}{\partial}

\newcommand{\nablabf}{\boldsymbol{\nabla}}

\renewcommand{\etal}{\textit{et~al.}}

\newcommand*{\plimsoll}{{\ensuremath{-\kern-4pt{\ominus}\kern-4pt-}}}
\newcommand{\scap}{\!\cdot\!}

\newcommand{\aaa}{\vec{a}}

\newcommand{\cfl}{c_\mr{fl}}

\newcommand{\EEE}{\vec{E}}

\newcommand{\eee}{\vec{e}}

\newcommand{\JJJ}{\vec{J}}

\newcommand{\rrr}{\vec{r}}

\newcommand{\sss}{\vec{s}}

\newcommand{\uuu}{\vec{u}}

\newcommand{\vvv}{\vec{v}}

\newcommand{\zerovec}{\boldsymbol{0}}

\newcommand{\III}{\boldsymbol{I}}
\newcommand{\kB}{k_\mathrm{B}}

\newcommand{\kapfl}{\kappa_\mathrm{fl}}

\newcommand{\eps}{\epsilon}

\newcommand{\etafl}{\eta_\mr{fl}}

\newcommand{\etab}{\eta_\mathrm{fl}^\mathrm{b}}

\newcommand{\Gamfl}{\Gamma_\mathrm{fl}}

\newcommand{\rhoel}{\rho_\mathrm{el}}
\newcommand{\rhofl}{\rho_\mathrm{fl}}

\newcommand{\SICel}{^\circ\!\textrm{C}}

\newcommand{\SIum}{\upmu\textrm{m}}

\newcommand{\SImm}{\textrm{mm}}
\newcommand{\SImum}{\textrm{\textmu{}m}}

\newcommand{\SImV}{\textrm{mV}}

\newcommand{\nn}{\nonumber}
\newcommand{\beq}[1]{\begin{equation} \eqlab{#1}}
\newcommand{\eeq}{\end{equation}}
\newcommand{\bsub}{\begin{subequations}}
\newcommand{\esub}{\end{subequations}}
\def\bal#1\eal{\begin{align}#1\end{align}}
\def\balat#1#2\ealat{\begin{alignat}{#1} #2 \end{alignat}}
\def\bsublab#1#2\esublab{\bsub \eqlab{#1} #2 \esub}
\def\bsubal#1#2\esubal{\bsublab{#1}\begin{align}#2\end{align} \esublab}
\def\bsubalat#1#2#3\esubalat{\bsublab{#1} \begin{alignat}{#2} #3 \end{alignat} \esublab}

\newcommand{\eqlab}[1]{\label{eq:#1}}
\renewcommand{\eqref}[1]{Eq.~(\ref{eq:#1})}
\newcommand{\eqnoref}[1]{(\ref{eq:#1})}

\newcommand{\eqsref}[2]{Eqs.~(\ref{eq:#1}) and~(\ref{eq:#2})}
\newcommand{\eqsnoref}[2]{(\ref{eq:#1}) and~(\ref{eq:#2})}

\newcommand{\figref}[1]{Fig.~\ref{fig:#1}}

\newcommand{\figlab}[1]{\label{fig:#1}}

\newcommand{\secref}[1]{Section~\ref{sec:#1}}

\newcommand{\seclab}[1]{\label{sec:#1}}
\newcommand{\tabref}[1]{Table~\ref{tab:#1}}

\newcommand{\tablab}[1]{\label{tab:#1}}

\newcommand{\grad}{\boldsymbol{\nabla}}

\renewcommand{\div}{\nablabf\!\cdot}

\newcommand{\lap}{\nabla^2}

\newcommand{\feo}{f_\mathrm{eo}}			
\newcommand{\fac}{f_\mathrm{ac}}			
\newcommand{\Veo}{V_\mathrm{eo}}			
\newcommand{\Vac}{V_\mathrm{ac}}			
\newcommand{\omgac}{\omega_\mathrm{ac}}	
\newcommand{\omgeo}{\omega_\mathrm{eo}}		
\newcommand{\omgeoIopt}{\omega_\mathrm{1,eo}^\mr{opt}}		 
\newcommand{\omgeonopt}{\omega_\mathrm{n,eo}^\mr{opt}}		 
\newcommand{\lambdaD}{\lambda_\mathrm{D}}	
\newcommand{\VT}{V_\mathrm{T}}			
\newcommand{\omgD}{\omega_\mathrm{D}}		
\newcommand{\phib}{\phi_\mathrm{1,fl}^\mathrm{bk}}		 
\newcommand{\phibn}{\phi_\mathrm{1,fl}^{\mathrm{bk},n}}		 
\newcommand{\phil}{\phi_\mathrm{1,fl}^{\lambda_\mathrm{D}}}	 
\newcommand{\vEO}{v^\mathrm{eo}}			
\newcommand{\Tac}{T_\mathrm{ac}}			
\newcommand{\Teo}{T_\mathrm{eo}}			
\newcommand{\epsfl}{\epsilon_\mathrm{fl}}		
\newcommand{\phifl}{\phi_\mathrm{fl}}		
\newcommand{\phiflI}{\phi_\mathrm{1,fl}}		
\newcommand{\phisl}{\phi_\mathrm{sl}}		
\newcommand{\epssl}{\epsilon_\mathrm{sl}}		
\newcommand{\vvvTA}{\avr{\vvv_2}}		
\newcommand{\vvvIIac}{\avr{\vvv^\mr{ac}_2}}		
\newcommand{\vvvIIeo}{\avr{\vvv^\mr{eo}_2}}		
\newcommand{\vIIac}{\avr{v^\mr{ac}_2}}		
\newcommand{\vvvSlipAC}{\avr{\vvv_\mathrm{2,slip}^\mathrm{ac}}}		 
\newcommand{\vvvSlipEO}{\avr{\vvv_\mathrm{2,slip}^\mathrm{eo}}}		 
\newcommand{\omgn}{\omega_n}
\newcommand{\omgm}{\omega_{n+1}}
\newcommand{\pIIac}{\langle p_2^\mathrm{ac}\rangle}
\newcommand{\pIIeo}{\langle p_2^\mathrm{eo}\rangle}
\newcommand{\nuI}{\nu_1}

\begin{document}

\title{Theory and simulation of AC electroosmotic suppression of acoustic streaming}

\author{Bjørn G. Winckelmann}
\email{winckel@dtu.dk}
\affiliation{Department of Physics, Technical University of Denmark,\\ DTU Physics Building 309, DK-2800 Kongens Lyngby, Denmark}

\author{Henrik Bruus}
\email{bruus@fysik.dtu.dk}
\affiliation{Department of Physics, Technical University of Denmark,\\
DTU Physics Building 309, DK-2800 Kongens Lyngby, Denmark}

\date{28 January 2021}

\begin{abstract}
Acoustic handling of nanoparticles in resonating acoustofluidic devices is often impeded by the presence of acoustic streaming. For micrometer-sized acoustic chambers, this acoustic streaming is typically driven from the fluid-solid interface by viscous shear-stresses generated by the acoustic actuation. AC electroosmosis is another boundary-driven streaming phenomena routinely used in microfluidic devices for handling of particle suspensions in electrolytes. Here, we study how streaming can be suppressed by combining ultrasound acoustics and AC electroosmosis. Based on a theoretical analysis of the electrokinetic problem, we are able to compute numerically a form of the electrical potential at the fluid-solid interface, which is suitable for suppressing a typical acoustic streaming pattern associated with a standing acoustic half-wave. In the linear regime, we even derive an analytical expression for the electroosmotic slip velocity at the fluid-solid interface, and use this as a guiding principle for developing models in the experimentally more relevant nonlinear regime that occurs at elevated driving voltages. We present simulation results for an acoustofluidic device, showing how implementing a suitable AC electroosmosis results in a suppression of the resulting streaming in the bulk of the device by two orders of magnitude.
\end{abstract}




\maketitle


\section{Introduction}
\seclab{intro}
Acoustofluidics is a rapidly advancing field of research based on the integration of ultrasound and microfluidics in lab-on-a-chip designs. Acoustic waves are used for label-free and efficient particle handling with high bio-compatibility, and the principle has found many application within biotechnology and health care. Examples include acoustic separation, \cite{Petersson2007, Manneberg2008, Ding2014} trapping, \cite{Hammarstrom2012, Evander2015} tweezing \cite{Collins2016, Lim2016, Baresch2016} as well as enrichment of cancer cells \cite{Augustsson2012, Antfolk2015}, and bacteria \cite{Antfolk2014, Li2017} and size-independent sorting of cells.\cite{Augustsson2016}

System designs for particle migration by ultrasound, termed acoustophoresis, are typically based on elongated fluid channels with cross  section dimensions in the range of 0.1 to 1~mm. The frequency of the acoustic waves in aqueous suspensions with wavelengths comparable to the chamber dimensions is thus in the low MHz range. These ultrasound fields are generated by piezoelectric transducers. Two competing forces of nonlinear origin act on particles suspended in the fluid. One force is the acoustic radiation force induced by acoustic wave-scattering by the particles.\cite{King1934, Yosioka1955, Doinikov1997, Settnes2012, Karlsen2015} This force focuses particles in nodes or antinodes of the acoustic waves, and it scales with the particle volume.\cite{Muller2013} The other force is the viscous Stokes drag due to the acoustic streaming,\cite{LordRayleigh1884, Schlichting1932, Westervelt1953, Nyborg1958} which scales linearly with the particle radius and tends to swirl particles around. Because of the different scalings, streaming-induced drag force is the dominating force for particles smaller than a critical size. For an aqueous suspension of spherical polystyrene particles in a 1-MHz ultrasound field, the critical diameter has been determined to be around $2~\SIum$.\cite{Muller2012, Barnkob2012a} To ease the acoustic manipulation of sub-$\SImum$ particles, such as bacteria, viruses, and exosomes, we seek to suppress the acoustic streaming.

There are two types of acoustic streaming: The boundary-layer-driven streaming originating from the viscous boundary layers near the fluid-solid interfaces, as first analyzed analytically by Lord Rayleigh \cite{LordRayleigh1884} and later studied in more detail,\cite{Nyborg1958, Lee1989, Vanneste2011, Bach2018} and the bulk-driven streaming generated by attenuation of acoustic waves in the bulk of the fluid,\cite{Eckart1948} an effect that is typically negligible in microfluidics except for rotating acoustic fields.\cite{bach2019}

Electroosmosis, the steady motion of electrolytic solutions with respect to a charged surface by an external electric potential, is another type of boundary-driven streaming.\cite{Squires2004} The principle has been used to create, say, micropumps with no moving parts for lab-on-a-chip systems.\cite{Ajdari2000, Brown2000, Gregersen2007} In particular, AC electroosmotic pumps have gained attention by generating relevant flow velocities at relatively low ac-voltages without electrolysis, thus circumventing the problem of gas formation. Pumping velocities have been reported in the $\sim100\,\SImum/$s range,\cite{Brown2000, Gregersen2007} which is of similar magnitude to typical acoustic streaming velocities.

In this study, we suggest to combine acoustic and AC electroosmotic streaming with a resulting net streaming close to zero. As an example of how to achieve this, we propose a specific design of an electro-acoustic device consisting of a microchannel with surface electrodes for generating electroosmosis, embedded in an elastic solid with an attached piezoelectric transducer for generating acoustics. We emphasize that the analysis is carried out with no intrinsic zeta potential. The ensuing study thus belongs to the body of work, which assumes that the chemically generated intrinsic zeta potential has been removed, say by means of a DC offset in the applied potential or by chemical surface treatments. A thorough study of the effects of an intrinsic zeta potential is left for future work.

The paper is structured as follows:
In \secref{GovEqu} we present the general theoretical framework.
In \secref{Linear_analysis} we consider AC electroosmosis at low excitation voltages, where the electrokinetic problem can be linearized and analytical expressions for the electroosmotic slip velocity are obtained. In \secref{Non_lin_EO}, the analytical solution is compared to full numerical solutions at higher voltages, and we find that the linearized regime capture both qualitative and quantitative behavior at surprisingly high voltages. In \secref{2D_devices_sim}, the linearized AC electroosmotic theory is integrated into an existing acoustofluidic simulation adapted from Skov \etal\cite{Skov2019} Using an electrode design, which could be produced by standard clean room fabrication techniques, together with an attached piezoelectric transducer, we demonstrate how to obtain heavily suppressed streaming patterns when exciting both AC electroosmosis and ultrasound acoustophoresis. In \secref{Discussion}, we discuss some of the limitations of the proposed model, and finally we conclude.

\section{Theory}
\seclab{GovEqu}
The core part of the system we analyze is a binary electrolyte in the form of a dilute aqueous suspension of ions,
with a single positive ion (subscript $+$) and a single negative ion (subscript $-$) of opposite valences $Z_\pm = \pm Z$. The electrolyte is placed inside a microchannel in the presence of a MHz-ultrasound pressure resonance $p_1$ with angular frequency $\omgac = 2\pi\fac = 2\pi\frac{1}{\Tac}$ and an kHz-AC electrical potential $\phi$ with angular frequency $\omgeo = 2\pi\feo = 2\pi\frac{1}{\Teo}$. The fundamental continuum fields in the fluid at point $\rrr$ and time $t$ is the mass density $\rho$, the pressure $p$, the fluid velocity $\vvv$, and the viscous stress tensor  $\vec{\sigma}$, as well as the concentration fields of the positive and negative ions $c_+$ and $c_-$, the electrical charge density $\rhoel$, and the electric potential $\phi(\rrr,t)$. The material parameters of the system are the dynamic viscosity $\etafl$, the bulk viscosity $\etab$, the speed of sound $\cfl$, the quiescent mass density $\rhofl$, the isentropic compressibility $\kapfl$, the ionic valence number $Z$, the ionic diffusivities $D_\pm$ and mobilities $\mu_\pm$, as well as the electric permittivity $\epsfl$. The governing equations for the fluid including acoustics are the mass continuity and Navier--Stokes equation,
 \bsubal{fluid_gov}
 \eqlab{cont}
 \pp_t \rho&=-\div (\rho\vvv), \\
 \eqlab{NS}
 \pp_t(\rho\vvv)&=-\div\big[(\rho \vvv)\vvv\big]+\div \vec{\sigma}-Ze(c_+-c_-)\nablabf \phi,
 \\
 \vec{\sigma}&=-p\III\!+\!\etafl
 \big[\nablabf \vvv\!+\!(\nablabf\vvv)^T\big]+\tfrac{3\etab-2\etafl}{3}\nablabf\scap\vvv\: \III.
 \esubal
The electrokinetics of the ions are governed by the concentration continuity, Nernst, and Poisson equation,
\bsubal{electrokinetic_gov_intro}
\eqlab{Nernst_Planck_intro}
\pp_t c_\pm&=-\div \JJJ_\pm,\\
\eqlab{Flux_ions_intro}
\JJJ_\pm&=c_\pm \vvv-D_\pm \nablabf c_\pm-\mu_\pm c_\pm \vec{\nabla}\phi, \\
\eqlab{Poisson_intro}
\nabla^2\phi&=-\frac{1}{\epsfl}Ze(c_+-c_-).
\esubal

The fields $Q(\rrr,t)$ will be treated in perturbation theory written as
 \beq{Ffield}
 Q(\rrr,t) = Q_0(\rrr) + \re\Big[Q_1(\rrr)\,\ee^{-\iot}\Big] + Q_2(\rrr,t).
 \eeq
Here, $Q_0$ is the unperturbed field, $Q_1$ is the first-order acoustic and electric time-harmonic perturbation, and $Q_2$ is the unsteady second-order field, which is generated by the inherent nonlinearities in hydrodynamic and electrokinetic equations. The time-averaged second-order response is defined $\avr{Q_2(\rrr,t)} = \frac{\omega}{2\pi} \int_0^{\frac{2\pi}{\omega}} Q_2(\rrr,t)\, \dm t$. A time-average of a product of two first-order fields is also a second-order term, written as $\avr{A_1 B_1}=\frac{1}{2} \re \big[ A_1 B_1^*\big]$, where the asterisk denote complex conjugation.

\begin{table}[]
\centering
\caption{\tablab{param} Physical parameters at $25~\SICel$ used in the numerical simulation of the electroacoustic device: water, dilute KCl solutions, Pyrex glass, and the piezoelectric transducer Pz26.}
\begin{tabular}{lccc}
\hline
\hline
Parameter                    & Symbol                        & Value & Unit \\ \hline
\textbf{Water properties:}\cite{Muller2014} &                &       &
\\
Mass density                 & $\rho_\mr{fl}$            & $997$ & kg$\,\rm{m^{-3}}$
\\
Compressibility              & $\kappa_\mr{fl}$          & $447$ & $\rm{TPa^{-1}}$
\\
Speed of sound               & $c_\mr{fl}$               & $1497$& $\rm{m\,s^{-1}}$
\\
Shear viscosity              & $\eta_\mr{fl}$     &$8.900\times 10^{-4}$&$\rm{Pa\, s}$
\\
Bulk viscosity               & $\eta_\mr{fl}^\mr{b}$ &$2.485\times 10^{-3}$&$\rm{Pa\, s}$
\\
Electrolyte permittivity\cite{CRC}     & $\eps_\mr{fl}$ &$6.938\times 10^{-10}$&$\rm{F\,m^{-1}}$
\\
\textbf{Ion properties:}     &                               &       &
\\
Average diffusivity\cite{CRC} & $D$             & $1.995\times 10^{-9}$&$\rm{m^2\, s^{-1}}$
\\
Valence                      & $Z$                           & $1$   &   -
\\
Bulk concentration           & $c_0$                         & $1$   &  $\rm{mM}$
\\
Debye length                 & $\lambda_\mr{D}$          & $9.6$ & $\rm{nm}$
\\
Debye frequency              & $\omega_\mr{D}$&$2.16\times 10^{7}$&$\rm{rad\,s^{-1}}$
\\
\textbf{Pyrex properties:}\cite{CorningPyrex}   &                               &       &
\\
Mass density                 & $\rho_\mr{sl}$            & $2230$&$\rm{kg\,m^{-3}}$
\\
Permittivity\cite{CRC}   & $\eps_\mr{sl}$            & $4.073\times 10^{-11}$ &$\rm{F\,m^{-1}}$
\\
Longditudinal sound speed    & $c^\mr{Py}_\mr{L}$    & $5592$& $\rm{m\,s^{-1}}$
\\
Transverse sound speed	     & $c^\mr{Py}_\mr{T}$    & $3424$& $\rm{m\,s^{-1}}$
\\
\textbf{Pz26 properties:}\cite{Ferroperm2017}    &                               &       &
\\
Mass density                 & $\rho_\mr{Pz26}$          &$7700$ &$\rm{kg\,m^{-3}}$
\\
Elastic constants:   		 & $c_{11}$           &$1.68\times 10^{11}$&$\rm{Pa}$
\\
                             & $c_{12}$           &$1.10\times 10^{11}$       &$\rm{Pa}$
\\
                             & $c_{13}$           &$9.99\times 10^{10}$       &$\rm{Pa}$
\\
                             & $c_{33}$           &$1.23\times 10^{11}$       &$\rm{Pa}$
\\
                             & $c_{44}$           &$3.01\times 10^{10}$       &$\rm{Pa}$
\\
                             & $c_{66}$           &$2.90\times 10^{10}$       &$\rm{Pa}$
\\
Coupling constants: 		 & $e_{31}$           &$-2.80$       &   $\rm{C\,m^{-2}}$
\\
                             & $e_{33}$           &$14.70$       &  $\rm{C\,m^{-2}}$
                             \\
                             & $e_{15}$           &$9.86$       & $\rm{C\,m^{-2}}$
                             \\
Permittivities:    			 & $\epsilon_{11}$  &$7.3313\times 10^{-9}$& $\rm{F\,m^{-1}}$
\\
                             & $\epsilon_{33}$ & $6.1979\times 10^{-9}$&  $\rm{F\,m^{-1}}$    	
\\	
\hline
\hline
\end{tabular}
\end{table}

\subsection{Combined acoustics and AC electroosmosis}

\begin{figure}[t]
\centering
\includegraphics[width=\columnwidth]{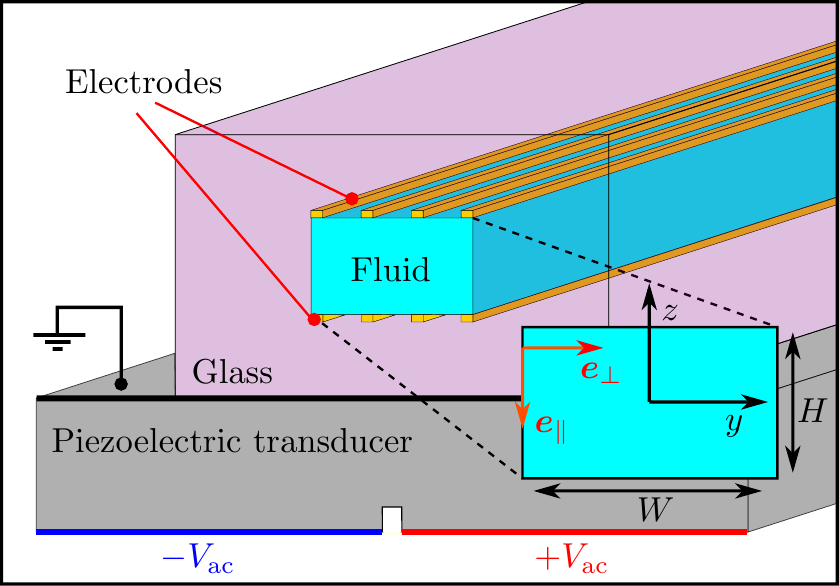}
\caption{\figlab{DeviceSketch} A sketch of the proposed electroacoustic device: a microchannel in glass with integrated acoustics and electroosmosis. Electrode arrays with AC-voltage $\Veo$ are implemented at the fluid-solid interface to induce the AC electroosmosis, and a piezoelectric transducer with AC-voltage $\Vac$ is glued to the bottom of the device to generate acoustics.}
\end{figure}

We consider a microfluidic system with integrated acoustics and electroosmosis. A conceptual sketch of such a system is shown in \figref{DeviceSketch}. A piezoelectric transducer actuates the system acoustically and generates acoustic streaming, and electrode arrays surrounding the fluid channel actuate AC electroosmotic streaming, which by proper design aims to counteract and suppress the acoustic streaming. To provide a proof of concept of this streaming suppression, we consider a simple long, straight rectangular fluid channel of dimensions $W\times H=375\,\SImum\times 160\,\SImum$. The acoustic problem in this configuration has previously been studied extensively both theoretically and experimentally,\cite{Muller2012, Muller2013, Muller2014, Muller2015} and it is known that the physical properties of the system are well-described by modeling restricted to the two-dimensional (2D) cross section. We thus apply a Cartesian ($y,z$) coordinate system centered in the fluid channel. The equilibrium position of the fluid-solid interface will be denoted $\sss_0$, and to describe boundary effects, we apply a local coordinate system $(\eee_\parallel,\eee_\perp)$ at the boundary.

Combining the two phenomena could potentially lead to non-trivial coupled effects. When acoustic waves travel through a ionic suspension, the ions will oscillate slightly out of phase with respect to the solvent. The different mobilities of ionic species will lead to a so-called ionic vibration potential. These potentials normalized by the oscillatory velocity of the fluid are typically of the order of $1\,\mr{mV}/(1\,\mr{m/s})$ \cite{Zana1982}. This effect is around two orders of magnitude lower than what is needed for acoustic streaming suppression by electroosmosis, and it is thus ignored.

As we shall see in the following, AC electroosmotic flows work ideally for $\feo\sim 1\,\mr{kHz}$, whereas the acoustic actuation frequencies are 1000 times faster in the range of $\fac\sim 1\,\mr{MHz}$. This separation of time scales allows us to use the acoustomechanical responses at the time-averaged (with respect to $T_\mr{ac}$) spatial position of the fluid-solid interface and the oscillating fluid, when computing the electrokinetic responses. Furthermore, the electrokinetic flow is established through the ionic Debye layer at the fluid-solid boundary on the short length scale $\lambdaD\sim 10\,\mr{nm}$, whereas the acoustically induced velocity fields are established over a viscous boundary layer of the much longer length scale $\delta\sim 500\,\mr{nm}$. Thus, we assume that no significant advection of ions in the Debye layer happen due to the acoustic streaming or vibrational velocity. This spatiotemporal decoupling of the electrokinetics and acoustics is further supported by the fact, that both electroosmotic and acoustic streaming in microchannels are described by linear Stokes flows,\cite{Ajdari2000, Bach2018} so we conclude that the combined streaming can be derived simply by superimposing the two flows computed separately.

Lastly, as the electric field extends throughout the fluid, dielectrophoretic forces inevitably arise and act on suspended particles.\citep{Voldman2006} For most materials in the present context, these forces are several orders of magnitude lower than the acoustic radiation force and the drag forces from streaming. We therefore ignore dielectrophoresis is this analysis.

\subsection{Pressure acoustics with viscous boundary layers}
To simulate acoustic fields, we follow the approach of Refs.~\onlinecite{Bach2018,Skov2019}.
The complex amplitude of the mechanical displacement field in the surrounding solid of the fluid channel and in the piezoelectric transducer is denoted $\uuu_1$. The complex acoustic pressure field and the associated oscillating fluid velocity field are denoted $p_1$ and $\vvv_1$, respectively. The steady time-averaged acoustic streaming $\vvvIIac$ and the corresponding pressure $\pIIac$ are then calculated as a Stokes flow with an acoustic slip velocity $\vvvSlipAC$ at the boundary,
 \bsubal{v2_gov}
 \eqlab{v2_gov_cont}
 0 &= \div \vvvIIac ,
 \\
 \eqlab{v2_gov_navier}
 \zerovec &= -\grad \pIIac
 + \etafl\lap \vvvIIac
 +\frac{\Gamfl\omgac}{2\cfl^2}\re[p_1\vvv_1],
 \\
 \eqlab{v2_gov_bc}
 \vvvIIac &=\vvvSlipAC,\; \text{ for } \rrr = \sss_0.
 \esubal
In \eqref{v2_gov_navier}, the acoustic body force, with the small viscous bulk damping coefficient  $\Gamfl=\big[\frac43\etafl+\etab\big]\kapfl\omgac\ll 1$, is typically negligible for single mode operation in microchannels, \cite{bach2019} and the flow is thus mostly driven by the slip velocity $\vvvSlipAC$.
We consider the conventional standing half-wave mode in the acoustic pressure. As we demonstrate later, the acoustic slip velocity for this mode closely resemble that of the classical Rayleigh streaming in fluid channels etched into acoustically hard materials like pyrex glass,
 \bsubal{v2_ray}
 \eqlab{v2_rayleigh}
 \vvvSlipAC_{z=\pm \frac12 H} &\approx \eee_y v_2^\mr{Rayl}\sin\!\big(2k_0 y\big), \\
 \eqlab{v2_rayleigh_amp}
 v_2^\mr{Rayl}&=\frac{3E_\mr{ac}}{2\rhofl \cfl},
 \esubal
where $E_\mr{ac}$ is the average acoustic energy density,
 \beq{EacDef}
 E_\mr{\mr{ac}} = \int_{-\frac{W}2}^{\frac{W}2}\int_{-\frac{H}2}^{\frac{W}2}
 \bigg[\frac{1}{4}\rhofl |\vvv_1|^2+\frac{1}{4}\kapfl |p_1|^2\bigg] \, \frac{\dm y \dm z}{HW}.
 \eeq

To determine the suppression of streaming numerically, we seek to minimize the average steady streaming in the fluid cross section,
 \beq{Avg_v2}
 \big|\avr{\vvv}\big|_\mr{avg} =
 \int_{-\frac{W}2}^{\frac{W}2}\int_{-\frac{H}2}^{\frac{H}2}
 \big|\avr{\vvv(y,z)}\big|\, \frac{\dm y \dm z}{HW}
 \eeq
Inspired by Ref.~\onlinecite{Bach2020}, the streaming suppression is also quantified by the measure,
 \bsubal{Suppression_parameter}
 \eqlab{Sup_param}
 S_q &= \int_{-\frac{W}2}^{\frac{W}2}\int_{-\frac{H}2}^{\frac{H}2}
 \Theta\Big(\frac{q}{100}v_2^\mr{Rayl}-|\avr{\vvv}|\Big)\, \frac{\dm y \dm z}{HW}, \\
 \eqlab{Heaviside}
 \Theta(x) &= \begin{cases}
      0, & x< 0, \\
      1, & x\geq 0.
   \end{cases}
 \esubal

\begin{figure}[t]
\centering
\includegraphics[width=\columnwidth]{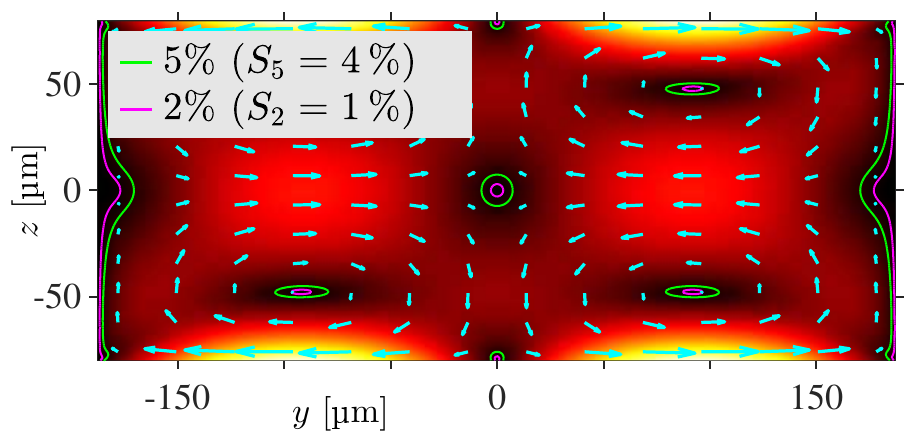}
\caption{\figlab{Acoustic_mode} Vector plot (cyan) and color plot from 0 (black) to $v_2^\mr{Rayl}$ (white) of the simulated acoustic streaming $\vvvIIac$ without electrokinetics generated by the side-wall actuation \eqref{acoustic_act} in the rectangular cross section $W \times H = 375~\SImum \times 160~\SImum$. $2\,\%$ (magenta) and $5\,\%$ (green) velocity contours of $\vIIac/v_2^\mr{Rayl}$ are shown with the suppression values $S_2$ and $S_5$, see \eqref{Sup_param}.}
\end{figure}

For the initial part of our study, we employ the analytically known acoustic resonance mode derived in Ref.~\onlinecite{Bach2018}, where the side walls are oscillated as,
\bal
\eqlab{acoustic_act}
\uuu_1(\pm \tfrac12 W,z)=d_0 \cos\!\bigg(\sqrt{-(1+\ii)\frac{\delta}{H}}\frac{\omgac}{\cfl}z\bigg)\, \ee^{-\ii\omgac t}.
\eal
Here, $d_0$ is the displacement amplitude, which is tuned to reach a desired average energy density or acoustic streaming. Given the physical parameters used for our study and listed in table \tabref{param}, the resonance frequency for water at $25~\SICel$ is $f_\mr{ac}^\mr{res}=1.993$~MHz. The acoustic streaming pattern generated by this actuation is shown in \figref{Acoustic_mode}.

\subsection{AC electroosmosis}
We consider an aqueous solution of a simple binary salt, say KCl, with ionic charges $Z_\pm = \pm Z$, concentrations $c_\pm$, diffusivities $D_\pm$, and electric mobilities $\mu_\pm$. We also introduce the average diffusivity $D=\frac12(D_+ + D_-)$, which turns out to be the lowest order correction to asymmetric ions in the linearized theory given below. All parameter values are listed in \tabref{param}.

We largely follow the presentations in Refs.~\onlinecite{Ajdari2000, Mortensen2005} but consider a more general externally applied electric potential at the fluid-solid interface,
\bal
\eqlab{time_dep_pot_bc}
V_\mr{ext}(\sss_0,t)=\re\!\big[\Veo\:w(\sss_0)\:\ee^{-\ii\omgeo t}\big].
\eal
Here, $w(\sss_0)$ is some complex-valued function of order unity that describes the shape of the externally applied potential at the boundary, whereas $\Veo$ describes its amplitude. We consider flow velocities $\vvv^\mr{eo}$ of sufficiently low amplitudes to describe the fluid as incompressible and drop the nonlinear term in the momentum equation, \eqref{NS}.  The electric potential is denoted $\phifl$ in the fluid and $\phisl$ in the surrounding solid and piezoceramic. The full set of governing equations for the electrokinetic problem in the fluid are thus written as,
 \bsubal{electrokinetic_gov}
 \eqlab{Nernst_Planck}
 \pp_t c_\pm&=-\div \JJJ_\pm,\;  \text{ with }\; \div \vvv^\mr{eo} = 0,
 \\
 \eqlab{Flux_ions}
 \JJJ_\pm&=c_\pm \vvv^\mr{eo}-D_\pm \nablabf c_\pm-\mu_\pm c_\pm \vec{\nabla}\phifl,
 \\
 \eqlab{Poisson}
 \nabla^2\phifl&=-\frac{1}{\epsfl}Ze(c_+-c_-),
 \\
 \eqlab{Stokes}
 \rhofl \pp_t   \vvv^\mr{eo}
 &= -\nablabf p^\mr{eo}+\etafl\nabla^2 \vvv^\mr{eo}-Ze(c_+-c_-)\nablabf \phifl.
 \esubal

The boundary conditions for the electrokinetic problem on electrode surfaces $\pp \Omega_\mr{eo}$ is an equipotential condition and on dielectric surfaces $\pp \Omega_\mr{di}$ a zero-charge condition. All fluid-solid interfaces admit zero ionic flux normal to the wall (we adopt the notation $\pp_\perp=\eee_\perp\cdot \nablabf$ and $\pp_\parallel=\eee_\parallel\cdot \nablabf$), and the fluid velocity field must fulfill the no slip boundary condition,
 \bsubal{electrokinetic_bc}
 \eqlab{wall_pot}
 \phifl(\sss_0,t)&=V_\mr{ext}(\sss_0,t),& &\hspace{-0mm}\sss_0\in\pp \Omega_\mr{eo},
 \\
 \eqlab{dielec_bc}
 \epsfl\pp_\perp\phifl &=  \epssl\pp_\perp\phisl,& &\hspace{-0mm}\sss_0\in\pp \Omega_\mr{di},
 \\
 \eqlab{no_slip}
 \vvv^\mr{eo}(\sss_0,t)&=\vec{0},
 \\
 \eqlab{no_normal_flux}
 \eee_\perp\cdot \JJJ_\pm(\sss_0,t) &= 0.
 \esubal

\section{Linearized analysis of AC electroosmosis}
\seclab{Linear_analysis}
The nonlinear nature of \eqref{electrokinetic_gov} makes it computationally and analytically challenging to work with general electrokinetic problems. For most computations we opt to use a linearized theory, which heavily reduces the computational footprint. In \secref{Non_lin_EO} we address the error introduced by using this theory at higher voltages. A conceptual sketch of electroosmotic flow generation is shown in \figref{EO_conceptual}. The general idea is to generate an electric charge density at electrode surfaces and then inducing a boundary-driven flow by dragging the ions along the surface with a parallel electric field.

\subsection{Linearized electrokinetic equations}
\seclab{Lin_gov_eq_iceo}
We consider the special case where $D_\pm=D$ such that $\mu_+=-\mu_-=\mu$, because of the Einstein relation,
 \beq{mu_VT_def}
 \mu_\pm = \pm\frac{ZD_\pm}{\VT},\;\text{ with }\; \VT=\frac{\kB T}{e},
 \eeq
where $\kB$ is the Boltzmann constant, $T$ is the temperature, and $\VT \approx 26$~mV is the thermal voltage. Following the treatment in Ref.~\onlinecite{Mortensen2005}, we consider a zero intrinsic zeta potential and the linearized dynamical Debye--H{\"u}ckel regime, which is obtained for weak externally applied potentials $\Veo \ll \VT$ at the electrode surfaces.
Here, the applied potential and the corresponding changes in ionic densities act as first-order fields. Through the non-linear electric body force, this will generate steady streaming as well as double-harmonic streaming with frequency $2\omgeo$, similar to perturbative acoustic calculations. Denoting the initial ionic concentration by $c_0$, we write,
 \bsubal{phasor_elec}
 \eqlab{phasor_c}
 c_\pm &=  c_0+ \re\Big[c_{1\pm}\, \ee^{-\ii\omgeo t}\Big],
 \qquad |c_{1\pm}|\ll c_0, \\
 \eqlab{phasor_phi}
 \phifl &= \re\Big[\phiflI \, \ee^{-\ii \omgeo t}\Big],
 \quad
 \vvv^\mr{eo} =\vvv_2^\mr{eo}(\rrr,t),
 \quad
 p^\mr{eo} = p_2^\mr{eo}(\rrr,t).
 \esubal
The second-order electric field and ionic concentrations are omitted, as they do not affect the electroosmotic streaming.
We then apply the linearization $c_\pm\nablabf\phifl\approx c_0\nablabf\phiflI$, and $c_\pm \vvv^\mr{eo}\approx 0$ in \eqref{Flux_ions}, where the latter is valid in the diffusive limit, where ionic advection becomes insignificant compared to diffusion. The first-order electrodynamic fields and the steady time-averaged second-order flow are thus obtained from,\cite{Mortensen2005}
 \bsub
 \eqlab{electrokinetic_gov_lin}
 \bal
 \eqlab{Nernst_Planck_lin}
 -\ii\omgeo\: \nuI&=D\nabla^2 \nuI-\omgD\nuI,
 \\
 \eqlab{Poisson_lin}
 \nabla^2\phiflI&=-\frac{1}{\epsfl}Ze\nuI,
 \eal
 \bal
 \eqlab{Incom_lin}
 0&=\div \vvvIIeo,
 \\
 \eqlab{Stokes_lin}
 \vec{0}&=-\nablabf \pIIeo+\etafl\nabla^2 \vvvIIeo-\frac{Ze}{2} \re\Big[\nuI\nablabf \phiflI^*\Big],
 \eal
 \esub
where we have introduced the notation
\bal
\nuI=c_{1+}-c_{1-}, \quad\omgD=\frac{D}{\lambdaD^2}, \quad \lambdaD=\sqrt{\frac{\epsfl\kB T}{2(Ze)^2c_0}}.
\eal
Correspondingly, the boundary condition~\eqnoref{no_normal_flux} becomes,
 \beq{no_flux_lin}
 \eee_\perp \cdot \Big[\nablabf \nuI + \frac{\epsfl}{Ze\lambdaD^2}\nablabf\phiflI\Big] = 0,\;
 \text{ for }\; \rrr = \sss_0.
 \eeq

\begin{figure}[t]
\centering
\includegraphics[width=\columnwidth]{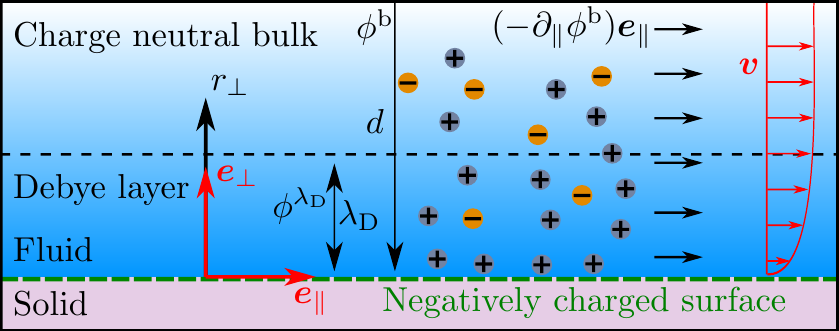}
\caption{\figlab{EO_conceptual} Conceptual drawing of electroosmosis at a flat boundary. Ions are pulled towards a charged surface, resulting in a thin layer of excess charge density termed the Debye layer. An electric field parallel to the surface is in turn established to pull on the charge density, which drives a flow through the electric body force present in the Debye layer.}
\end{figure}

\subsection{Effective AC electrokinetic theory}
\seclab{Electrokinetic_eff_bc}

The rectangular cross section $W\times H$ of \figref{Acoustic_mode} only contains planar fluid-solid interfaces, and the potential-shape function $w(\sss_0)$ of \eqref{time_dep_pot_bc} is assumed to vary on length scales comparable to the chamber dimensions $\pp_\parallel w(\sss_0)\sim \frac1d w(\sss_0)$, where $d\sim {W,H}$. Further, we assume that $\omgeo\ll\omgD$, which is the relevant limit for our purpose.
When an external potential is applied at the fluid-solid interface, ions will accumulate at the wall in a thin layer of length scale $\lambdaD$. This ionic layer will completely screen off the wall potential for DC wall potentials, but for AC potentials the screening is only partial. Because $\lambdaD \ll d$, we have $\nabla^2\sim \pp_\perp^2$ in \eqref{Nernst_Planck_lin}. The perpendicular coordinate away from the surface is called $r_\perp$, and thus
 \beq{nu_1st}
 \nuI(\rrr) = \nu_0\:w(\sss_0)\:\ee^{-\kappa r_\perp},
 \;\text{ with }\; \kappa=\frac{1}{\lambdaD}\sqrt{1-\ii \frac{\omgeo}{\omgD}},
 \eeq
where $\nu_0$ is a constant. The solution for $\phiflI$ will contain a particular solution $\phil$, which reflects the partial ionic screening in the thin Debye layer near the boundary, and a homogeneous solution $\phib$ that extends into the bulk,
 \beq{phi}
 \phiflI(\rrr) = \phil(\rrr) + \phib(\rrr).
 \eeq
Combining \eqsref{Poisson_lin}{nu_1st} and inserting $\nabla^2 \phil \sim \pp_\perp^2 \phil$, the two terms in \eqref{phi} obey,
 \bsubal{phi_calc}
 \eqlab{phi_lambda_calc}
 \phil(\rrr)&=-\nu_0\:\frac{Ze}{\epsfl \kappa^2}\:w(\sss_0)\:\ee^{-\kappa r_\perp},
 \\
 \eqlab{bulk_phi_calc}
 \nabla^2 \phib&=0.
 \esubal
Using boundary conditions \eqsnoref{wall_pot}{no_flux_lin} with Eqs.~\eqnoref{nu_1st}, \eqnoref{phi}, and \eqnoref{phi_lambda_calc} inserted at $\Omega_\mr{eo}$, we eliminate $\nu_0$ to find
 \bsubal{electro_eff}
 \eqlab{electro_eff_BC}
 \phib(\sss_0)&=\Veo w(\sss_0)+\frac{\ii}{\kappa}\frac{\omgD}{\omgeo}\pp_\perp \phib(\sss_0),\;
 \sss_0\in\Omega_\mr{eo},
 \\
 \eqlab{nu}
 \nuI(\rrr)&=\ii \frac{\omgD}{\omgeo}\frac{\epsfl \kappa}{Ze}\pp_\perp \phib(\sss_0)\ee^{-\kappa r_\perp},
 \\
 \eqlab{phi_lambda}
 \phil(\rrr)&=-\ii \frac{\omgD}{\omgeo}\frac{1}{\kappa}\pp_\perp \phib(\sss_0)\ee^{-\kappa r_\perp}.
 \esubal
For dielectric surfaces $\Omega_\mr{di}$, the boundary condition~\eqnoref{dielec_bc} can similarly be rewritten in terms of $\phib$,
 \beq{dielec_BC_lin}
 \epssl \pp_\perp\phisl(\sss_0) =\Big(1+\ii \frac{\omgD}{\omgeo}\Big)\epsfl \pp_\perp \phib(\sss_0),
 \; \sss_0\in\Omega_\mr{di}.
 \eeq
The known forms of $\nuI$ and $\phil$ can be inserted in \eqref{Stokes_lin} alongside the calculated $\phib$. With this, one can compute the steady time-averaged streaming $\vvvIIeo$ to lowest order in $\lambdaD/d$ and $\omgeo/\omgD$ as a simple Stokes flow with an electroosmotic slip velocity $\vvvSlipEO$ given by $\phib$ as,
 \bsubal{EO_eff}
 \eqlab{EO_eff_cont}
 0&=\div \vvvIIeo,
 \\
 \eqlab{EO_Stokes}
 \vec{0}&=\etafl\nabla^2\vvvIIeo-\nablabf \pIIeo,
 \\
 \eqlab{EO_slip_BC}
 \vvvIIeo_{\sss_0} &= \vvvSlipEO,
 \\
 \eqlab{EO_slip}
 \vvvSlipEO&=-\eee_\parallel \frac{\epsfl\omgD}{2 \etafl\omgeo}\re\!\Big[\frac{\ii}{\kappa}\pp_\perp\phib \big(\pp_\parallel\phib\big)^*\Big]_{\sss_0}.
 \esubal
This form of the slip velocity is essentially identical to the one given in Eq.~(2) of Ref.~\onlinecite{Ajdari2000}.

\subsection{The analytic electrokinetic double mode}
\seclab{Electrokinetic_double_mode}
To generate an AC electroosmotic slip velocity $\vvvSlipEO$ opposite to the acoustic Rayleigh slip velocity~\eqnoref{v2_ray}, we use an analytical model in the linearized regime as guidance. We consider an idealized case, where a perfectly smooth potential is generated at the fluid boundary, corresponding to the limit of implementing infinitely many infinitely thin electrodes in our electrode arrays. The more realistic case of a finite amount of electrodes is subsequently assessed in \secref{2D_devices_sim}.

In an experimental setup, see \figref{DeviceSketch}, it would likely be desirable to only implement electrodes at the top and bottom boundaries $(y,\pm\frac12 H)$ of the microchannel and thus assume a zero-charge conditions at the side walls $(\pm \frac12 W,z)$. The electric potential in the surrounding solid originate from the same electrodes that generate the fluid potential, which leads to $\phisl\sim\phib$. Because most relevant materials for creating microfluidic channels have $\epssl \ll \epsfl$, \eqref{dielec_BC_lin} dictates $\pp_\perp\phib \approx 0$ at a dielectric boundary.

A single sinusoidal mode in the electric potential is studied in Ref.~\onlinecite{Mortensen2005} and shown to give a negligible steady streaming component. Since $\kappa\approx 1/\lambdaD$ for $\omgeo\ll \omgD$, it is clear from \eqref{EO_slip} that a phase difference is needed between the perpendicular- and the parallel component of the electric bulk field $\EEE^\mr{bk}=-\nablabf \phib$ at the boundary $\sss_0$ to generate a significant steady streaming $\vvvIIeo$. This is not possible with a single mode, so instead we combine two sinusoidal modes with a relative phase difference $\vartheta$,
 \bsubal{BC_double_mode}
 \eqlab{BC_double_mode_sides}
 \pp_\perp \phibn(\pm \tfrac12 W,z) &= 0,
 \\
 \eqlab{BC_double_mode_top_bot}
 w(y,\pm \tfrac12 H) &= \sin(k_n y) + \ee^{\ii\vartheta}\sin(k_{n+1} y),
 \\
 \eqlab{knDef}
 \text{ with }\; k_n  & = \frac{2n+1}{W}\:\pi,\;\text{ for }\; n = 0, 1, 2, \ldots.
 \esubal
We use that $\kappa\approx 1/\lambdaD$, and derive the analytical solution of \eqsref{bulk_phi_calc}{electro_eff_BC} for the bulk potential $\phibn(y,z)$,
 \bsubal{phibkAnl}
 \phibn(y,z)&=\frac{\Veo\cosh\!\big(k_n z\big)\sin\!\big(k_n y\big)}{
 \cosh\!\big(k_n \frac{H}{2}\big)+\ii\frac{\omgn}{\omgeo}\sinh\!\big(k_n \frac{H}{2}\big)}
 \\ \nn
 &\quad +
 \frac{\ee^{\ii\vartheta}\Veo\cosh\!\big(k_{n+1} z\big)\sin\!\big(k_{n+1} y\big)}{
 \cosh\!\big(k_{n+1}\frac{H}2\big)+\ii\frac{\omgm}{\omgeo}\sinh\!\big(k_{n+1}\frac{H}2\big)},
 \\
 \text{with } \omgn  & = k_n \lambdaD \omgD,\;\text{ for }\; n = 0, 1, 2, \ldots.
 \esubal
Inserting this $\phibn$ in \eqref{EO_slip} leads to an expression for the electroosmotic slip velocity $\vvvSlipEO$, which depends on both $\vartheta$ and $\omgeo$. We are interested in generating large streaming amplitudes at low applied voltages. The phase difference that leads to the largest streaming amplitude is denoted $\vartheta = \vartheta^\mr{opt}_n$, and the optimized angular frequency at $\vartheta^\mr{opt}_n$ is denoted $\omgeo=\omgeonopt$. These can both be found analytically and result in the electroosmotic slip velocity,
\bsubal{slip_opt}
 \eqlab{slip_sides}
 &\vvvSlipEO|_{y=\pm \frac{W}{2}} = \vec{0},
 \\
 \eqlab{slip_top_bot}
 &\vvvSlipEO|_{z=\pm \frac{H}2} = -g_n(y)\:\vEO_n\: \eee_y
 \\ \nn
 &g_n(y) = \frac{\sinh(k_0 H)\sin(k_{2n+\frac32} y) \!+\!
 \sinh(k_{n+\frac12} \frac{H}2)\sin(2k_0 y)}{
 \frac{1}{2n+2}\sinh\!\big(k_0 H\big)+\sinh\!\big(k_{n+\frac12} \frac{H}2)},
 \\
 &\vEO_n = \frac{\pi \epsfl \Veo^2}{8\etafl W}\frac{(2n+1)(2n+3)}{n+1}.
 \esubal
The optimal phase difference $\vartheta^\mr{opt}_n$ and the corresponding optimal angular frequency $\omgeonopt$ are given by,
 \bsubal{opt_parameters}
 \eqlab{vartheta_opt}
 \vartheta^\mr{opt}_n &= \pi-\arctan(\alpha_n),
 \\
 \alpha_n &=
 \frac{1+\frac{\omgn\omgm}{\omgeo^2}\tanh\!\big(k_n \frac{H}{2}\big)\tanh\!\big(k_{n+1}\frac{H}2\big)}{
 \frac{\omgm}{\omgeo}\tanh\!\big(k_{n+1}\frac{H}2\big)-\frac{\omgn}{\omgeo}\tanh\!\big(k_n \frac{H}{2}\big)},
 \\
 \eqlab{omgel_opt}
 \omgeonopt &=
 \sqrt{\omgn\omgm\tanh(\tfrac12 k_n H)\tanh(\tfrac12 k_{n+1}H)}.
\esubal

The slip velocity $\vvvSlipEO$ in \eqref{slip_opt} approaches the desired acoustic slip velocity $-\vvvSlipAC$ of \eqref{v2_rayleigh} for large values of $n$. Even for $n=1$, the $\sin\!\big(\frac{2\pi y}{W}\big)$ term will dominate for $H\sim W$. The streaming amplitude is seen to increase linearly with $n$ for large values of $n$. This continuous growth is caused by the increase in the transverse electrical bulk field generated by the sinusoidal modes for a given surface potential amplitude.  In \figref{Ideal_suppression} is shown an electrokinetic simulation with the boundary conditions~\eqnoref{BC_double_mode} and $n=1$, superimposed on the acoustics simulation of \figref{Acoustic_mode}. The resulting suppressed streaming is shown next to the separate streaming patterns. For an average acoustic energy density of $E_\mr{ac}=100\,$Pa, the streaming is optimally suppressed through the electroosmotic mode discussed above at $\Veo=127\,$mV. We see that the streaming is suppressed below $5\,\%$ of the Rayleigh streaming amplitude everywhere in the channel, and almost the entire streaming pattern is suppressed below $2\,\%$. This result constitutes our first proof of concept of suppression of acoustic streaming by electroosmotic streaming.

Let us discuss the assumptions. It takes very high values of $n$ to violate the assumption $\lambdaD \ll d$, and using high order sinusoidal modes could be a valid strategy for creating powerful streaming. However, generating high order sinusoidal modes with discrete electrode arrays would require many electrodes, and in practice it may be desirable to use low order modes.
For $n=1$, we find $|\vvvSlipEO|_\mr{max}\approx 6\,$nm/s for $\Veo=1\,$mV. Extrapolating to higher voltages, one would need around $\Veo=130\,$mV to reach typical acoustic streaming velocities of around $100\,\SImum$/s. This obviously violates the assumption of $\Veo\ll \VT$, but still remains well below the steric regime.\cite{Bazant2007}

\begin{figure}[t]
\centering
\includegraphics[width=\columnwidth]{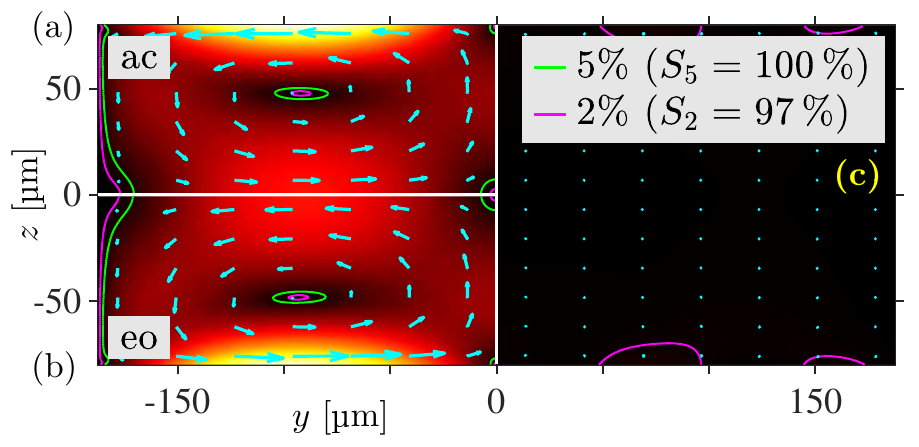}
\caption{\figlab{Ideal_suppression} Simulation results for the idealized model showing: (a) The acoustic streaming $\vvvIIac$ driven by the boundary condition in \eqref{acoustic_act}, (b) The electroosmotic streaming $\vvvIIeo$ generated by the boundary conditions in \eqref{BC_double_mode}. (c) The resulting strongly suppressed streaming $\vvvIIac+\vvvIIeo$ obtained for a typical acoustic energy density $E_\mr{ac}=100\,$Pa, combined with an electroosmotic streaming driven at $\Veo=127\,$mV.}
\end{figure}

The efficiency of the electric double-mode potential stems from a phase-matching between the charge density generated by one mode and the electric field of the other mode, which drags the established charge density along the top and bottom surfaces of the channel.

Lastly, we note that the suggested mode combination is not the only way of generating the desired streaming pattern. A combination between a sinusoidal and a linear mode with dielectric side-walls also generate a useful streaming pattern, albeit the analytical solution becomes more complicated. If one controls the potential on the side walls, it also becomes possible to generate clean sinusoidal modes with an even number of half-waves. These possibilities are not explored in this study.

\section{Non-linear electroosmotic flows}
\seclab{Non_lin_EO}
To reach experimentally relevant streaming velocities, one needs to apply voltages around $\Veo\sim 5\VT$, well into the nonlinear regime of the electrokinetic problem. In the following, we evaluate the capability of the linearized theory of AC electrokinetics to predict qualitative trends for higher voltages. We adopt a numerical approach inspired by Muller \etal\cite{Muller2015} to compare the linear predictions to the full nonlinear system of \eqref{electrokinetic_gov}. For lower voltages, where the linearized theory is valid, we assume the streaming velocity $\vvv^\mr{eo}$ and corresponding pressure $p^\mr{eo}$ to be of the form, \cite{Muller2015}
 \beq{NonLinForm}
 \aaa(\rrr,t) = \aaa_2^{0}(\rrr,t) + \aaa_2^{2\omgeo}(\rrr,t)\:\cos(2\omgeo t),
 \eeq
where the time dependencies in $\aaa_2^0$ and $\aaa_2^{2\omgeo}$ describe a transient period. For higher voltages, a more general time dependency could arise by significant mixing of Fourier components through the nonlinear terms in \eqsref{Flux_ions}{Stokes}. Using a fifth-order Romberg integration scheme,\cite{Muller2015, Press2002} the time-averaged response at a given time $t$ is computed numerically as,
 \beq{timeavrNum}
 \avr{\aaa(\rrr,t)} = \int_{t-\frac{1}{2}\Teo}^{t+\frac{1}{2}\Teo}\aaa(\rrr,t')\,\dm t'.
 \eeq

\subsection{Numerical implementation}
\seclab{non_lin_num_impl}
The governing equations~\eqnoref{electrokinetic_gov} are solved with the commercial finite element software COMSOL Multiphysics.\cite{Comsol54} Solving the full time-dependent nonlinear equations increases the computational footprint significantly compared to the linearized treatment.

We simulate the transient leading to the time-harmonic problem discussed in the linearized scheme above. The boundary condition~\eqnoref{BC_double_mode_top_bot} is used directly as formulated in \eqsref{time_dep_pot_bc}{wall_pot}. The boundary condition for the side-walls in \eqref{BC_double_mode_sides} is replaced by,
 \beq{numBC}
 \pp_\perp \phifl(\pm\tfrac12 W,z,t)=0.
 \eeq
The two central symmetry lines $x=0$ and $y=0$ were used to reduce the computational domain by a factor of four. This is illustrated in \figref{Ideal_suppression_non_lin}(a) alongside the corresponding symmetry boundary conditions. A structured mesh with high resolution near the domain boundaries was used to fully resolve the thin Debye layer, which now has to be taken into account numerically.

\begin{figure}[t]
\centering
\includegraphics[width=\columnwidth]{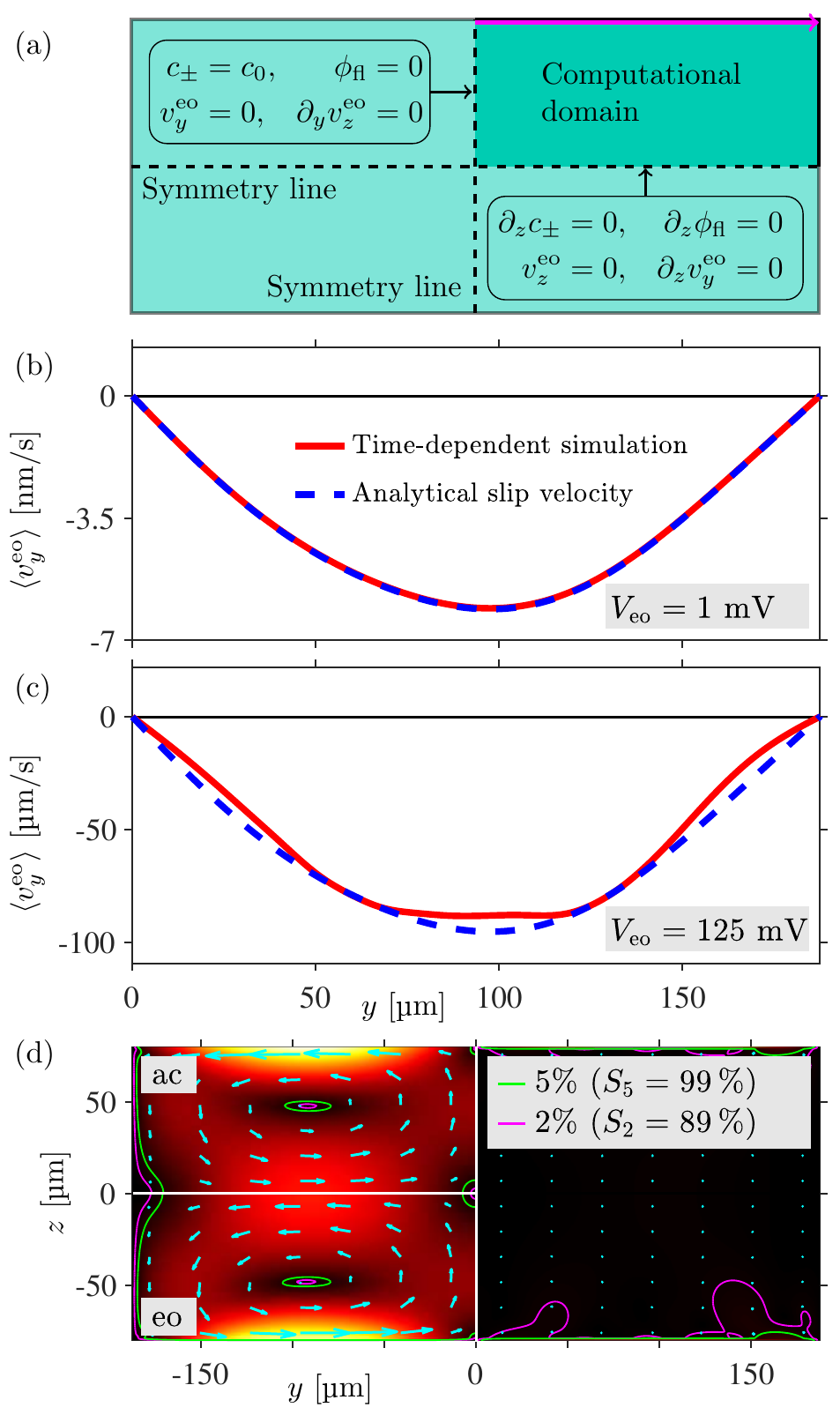}
\caption{\figlab{Ideal_suppression_non_lin} Geometry and results for the time-dependent, nonlinear, electroosmotic simulations. (a) solution domain with symmetry lines and symmetry boundary conditions illustrated. (b) comparison at $\Veo=1\,$mV between the slip velocity found in the time-dependent study and the linearized solution discussed above. The line plots are shown along the magenta arrow shown in (a). (c) corresponding comparison at $\Veo=125\,$mV. (d) field plot of the suppressed streaming pattern (right), and the two separate streaming patterns (left).}
\end{figure}

As in Ref.~\onlinecite{Muller2015}, a time-dependent solver is employed with the Generalized alpha time-stepping scheme, where the alpha factor is set to 0.5. A constant time-stepping of $\Delta t=\frac1{256}\Teo$ was used for all simulations. For simulations with $\Veo>1\,$mV, the system Jacobian was set to update at every time-step to stabilize the solutions.
The amplitude of the external voltage $\tilde{V}_\mr{eo}(t)$ was gradually ramped up through the first oscillation period as,
 \beq{Vramp}
 \tilde{V}_\mr{eo}(t) = \frac12 \big[1+\tanh\!\big(5\omgeo t-3\big)\big]\:\Veo.
 \eeq
The time-dependent solver was run for four electric periods, $0 < t < 4\Teo$, and as the physical system is not resonant, a steady state was reached already for $t \simeq 2\Teo$. Here, we present two different simulations, one linear with $\Veo = 0.04 \VT = 1~\SImV$ and the other nonlinear with $\Veo = 4.8~\VT = 125~\SImV$. The time-averaged solution of the latter, which took eight days to finish, turned out to contain higher-than-second-order harmonics with a relative amplitude of about 1\:\%, signalling the onset of the strongly nonlinear regime. However, due to their relatively low amplitude, these are not discussed further.

\subsection{Results of time-dependent simulations}
The simulation at $\Veo=1\,$mV was primarily made to check the numerical setup. In \figref{Ideal_suppression_non_lin}(b), the time-averaged horizontal velocity component $\avr{v_{y}^\mr{eo}}$ at $t=3.5\,\Teo$ is shown along the top right boundary, marked by the magenta arrow in \figref{Ideal_suppression_non_lin}(a). Specifically, the field is shown just outside the Debye layer at coordinates $(y,z)=(y,\frac{1}{2}H-7\lambdaD)$, and the comparison with the analytical expression~\eqnoref{slip_top_bot} for the linearized slip velocity $\vvvSlipEO$ shows an almost perfect agreement.

A similar plot is made for $\Veo=125\,$mV in \figref{Ideal_suppression_non_lin}(c), and there a notable difference between the two theories appears. The peak amplitude of the time-dependent simulation is $88\,\SImum/$s, whereas the linearized theory predicts $95\,\SImum/$s, 8\% higher. We however notice, that clear quantitative and even decent qualitative features remain. To validate the numerical result of this more nonlinear simulation, the first half period was simulated for increasing mesh refinement, which yielded no significant changes to the solution of the individual time-steps.

Finally, in \figref{Ideal_suppression_non_lin}(d) the time-averaged nonlinear electroosmotic flow $\avr{\vvv^\mr{eo}}$ at $\Veo=125\,$mV is superposed with the idealized acoustic streaming $\vvvIIac$ at $E_\mr{ac}=91\,$Pa, in a case chosen to minimize the resulting streaming $\avr{\vvv} = \avr{\vvv^\mr{eo}} + \vvvIIac$. As for the linear case \figref{Ideal_suppression}(c), the nonlinear streaming in \figref{Ideal_suppression_non_lin}(d)  is seen to be heavily suppressed. However here, in contrast to the linearized streaming, small patches of streaming extend from the boundary, where the time-dependent solution is seen to differ the most from the linearized theory. Nevertheless, the streaming is less than 2\% of the Rayleigh value $v_2^\mr{Rayl}$ in 89\% of the domain.

This brief study of electroosmotic streaming in the moderately nonlinear regime suggests that the linearized theory captures the main features of the streaming to a satisfactory degree at the suggested voltage range from 0 to $125~\SImV$ for the initial study presented in this work. We therefore return to the computationally much simpler linearized model in the next section, where we extend our model to include the elastic channel walls, a small number of surface electrodes, and the piezoelectric transducer.

\section{Numerical 2D device simulations}
\seclab{2D_devices_sim}
As sketched in \figref{DeviceSketch}, a typical acoustophoretic device contains a microchannel embedded in an elastic solid which is glued onto a piezoelectric transducer. Consequently, the acoustic response including the streaming is degraded compared to the ideal hard-wall system studied above. Moreover, it is not possible in a real electroosmotic device to create and control a given continuous shape of the surface potential, and instead only a limited number of finite-sized electrodes may be fabricated. In the following, we study through 2D numerical simulations, using the linearized electrokinetic model, to which extend the introduction of these more realistic aspects of the model will diminish the ability to suppress the acoustic streaming by electroosmotic streaming.

\subsection{The design of the 2D device}
\seclab{2D_device_design}
\begin{figure}[tb]
\centering
\includegraphics[width=\columnwidth]{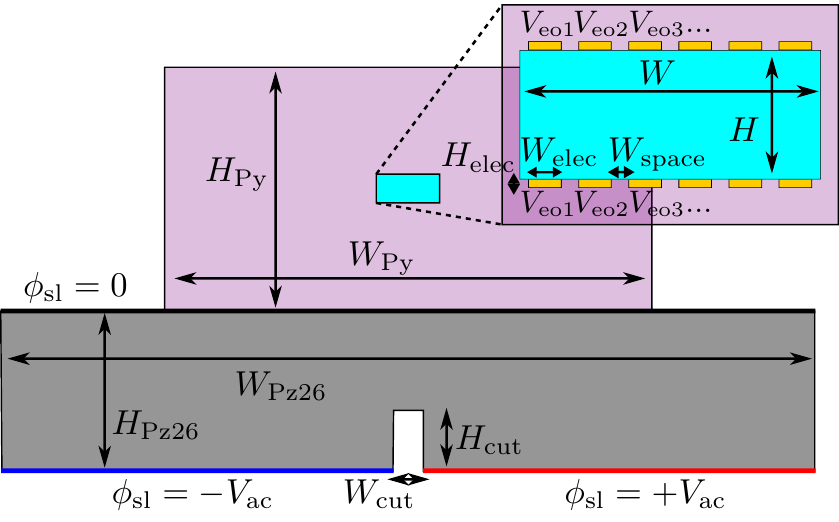}
\caption{\figlab{Simulation_geometry} Sketch of the 2D cross section of the simulated acousto-electrokinetic device consisting of a microchannel embedded in an elastic solid with surface electrodes and driven acoustically by an attached piezoelectric transducer with a split bottom-electrode for antisymmetric actuation. The relevant length scales and electrode potentials are labeled.}
\end{figure}

We consider the 2D model sketched in \figref{Simulation_geometry}. It contains a microchannel of dimensions $W\times H=375\,\SImum\times 160 \,\SImum$ filled with a dilute aqueous solution of KCl ions and embedded in an elastic block of pyrex glass of dimensions $H_\mr{Py}\times W_\mr{Py}=3\,\SImm\times 1.5 \, \SImm$. The electroosmotic streaming is actuated by voltages on the arrays of $N_\mr{elec}$ rectangular electrodes of dimensions $W_\mr{elec}\times H_\mr{elec}$ ($H_\mr{elec}=2\,\SImum$ and $W_\mr{elec}$ varied) and spacings  of dimension $W_\mr{space}$ engraved into the top and bottom fluid-solid interfaces. The acoustic streaming is actuated by the attached piezoelectric transducer of dimensions $W_\mr{Pz26}\times H_\mr{Pz26}=5\,\SImm\times 1\,\SImm$ modeled as the piezoelectric material Pz26 driven by the potential $\pm \Vac$. A central cut of dimensions $W_\mr{cut}\times H_\mr{cut}=160\,\SImum\times 375 \,\SImum$ is made in the bottom of the transducer to enable antisymmetric actuation.\cite{Bode2020}

To avoid excessive meshing in the numerical model near the electrodes at the fluid-solid interface, their thickness was chosen to be $2~\SImum$, which is about 40 times larger than standard clean-room-deposited electrodes typically having a thickness of $50\,\mr{nm}$. Since these enhanced electrodes still comprise a small fraction of the glass volume, they have a negligible effect on the resulting acoustic resonance properties of the channel. Acoustically the electrodes are part of the elastic solid, and electrically, they are modeled as ideal conductors having equipotential surfaces with the applied voltages $V_\mr{eo1},V_\mr{eo2}, \ldots, V_\mr{eo,N_\mr{elec}}$, respectively, where $V_1$ is applied to the outermost left electrodes at the top and bottom surfaces. The electrode- and spacing widths are always chosen such that $N_\mr{elec}(W_\mr{elec}+W_\mr{space})=W$ with $W_\mr{space}=\frac12 W_\mr{elec}$. The $y$-coordinates of the electrode centers $y_m$ for $m=1,2, \ldots, N_\mr{elec}$  and the potential $V_{\mr{eo,}m}$ applied to the electrode arrays, a discretized version of the potential shape~\eqnoref{BC_double_mode_top_bot} for  $n=1$, are then given by
 \bsubal{discretV}
 \eqlab{ymDef}
 y_m & =\frac{2m-1-N_\mr{elec}}{2N_\mr{elec}}\:W,\;\;
 m=1,2, \ldots, N_\mr{elec},
 \\
 \eqlab{VmDef}
 V_{\mr{eo,}m} &= \Veo\big[\sin(k_1 y_m)+\ee^{\ii\vartheta_1^\mr{opt}}\sin(k_2 y_m)\big].
 \esubal
The optimized parameters $\vartheta_1^\mr{opt}$ and $\omgeoIopt$ are used in all simulations. The side walls and electrode spacings will have a pyrex/water interface with the boundary condition described by \eqref{dielec_BC_lin}.

The system is actuated acoustically by potentials applied to the top and bottom boundaries of the piezoelectric transducer at a numerically determined resonance frequency $\fac^\mr{res}$. A voltage difference of $2\Vac$ is applied between the two bottom electrodes on the transducer, while the top electrode is grounded. $\Vac$ is chosen to reach an average acoustic energy density of $E_\mr{ac}=100\,$Pa.

\begin{figure}[b]
\centering
\includegraphics[width=0.8\columnwidth]{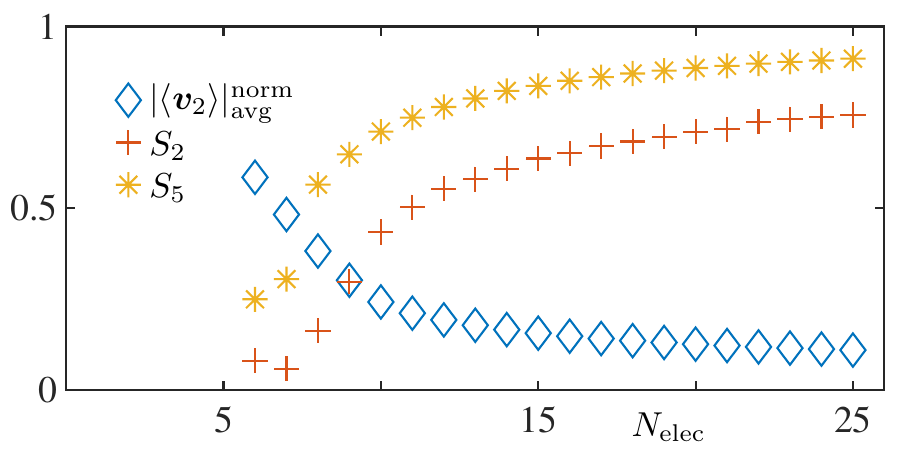}
\caption{\figlab{Nelec_sweep} Plots of normalized responses. Quantitative measures $|\langle \vvv_2\rangle|_\mr{avg}^\mr{norm}$, $S_2$, and $S_5$ of streaming suppression for the 2D system. $|\langle \vvv_2\rangle|_\mr{avg}^\mr{norm}$ is the average streaming amplitude from \eqref{Avg_v2} normalized with respect to the number found for $N_\mathrm{elec}=0$.}
\end{figure}

\begin{figure*}[t]
\centering
\includegraphics{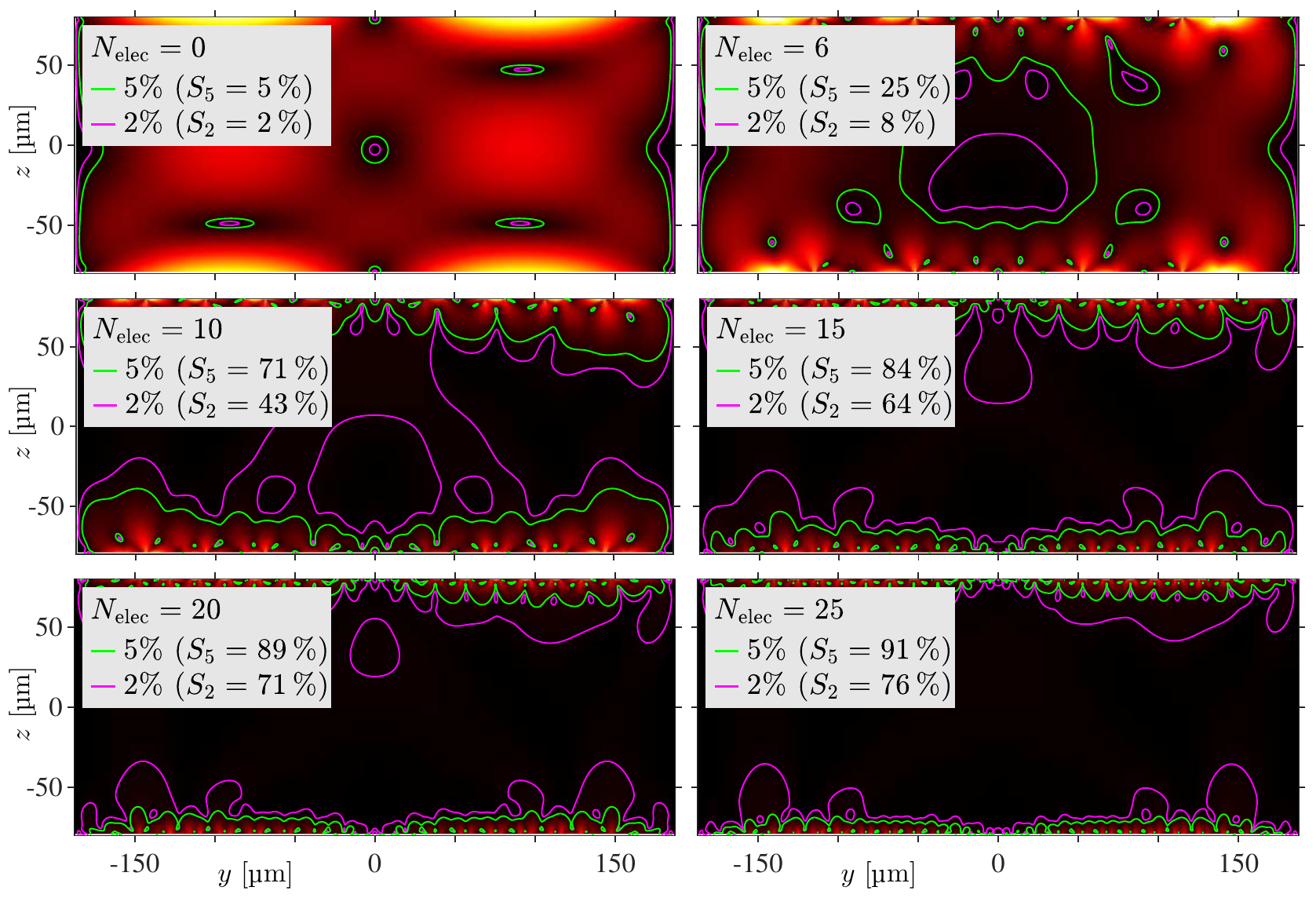}
\caption{\figlab{Discrete_electrodes} Field plots of the suppressed streaming patterns generated by the acoustophoretic system with integrated discrete electrodes for electroosmosis. Optimized solutions for increasing electrode counts are shown. Streaming arrows are omitted here for visual clarity.}
\end{figure*}

\subsection{Simulation results for the 2D device}
\seclab{2D_device_results}

In the following, we simulate the combined acoustic and electroosmotic streaming as the Stokes flow~\eqnoref{v2_gov} with its acoustic body force and acoustic slip, but now also adding the electrokinetics~\eqnoref{bulk_phi_calc} including the  electroosmotic slip velocity~\eqnoref{EO_slip},
 \bsub
 \eqlab{v2tot_gov}
  \bal
 \eqlab{v2tot_gov_cont}
 0 &= \div\vvvTA,
 \\
 \eqlab{v2tot_gov_navier}
 \zerovec &= -\grad \langle p_2 \rangle
 + \etafl\lap\vvvTA
 +\frac{\Gamfl\omgac}{2\cfl^2}\re[p_1\vvv_1],
 \eal
 \bal
 \eqlab{Poisson_lin_tot}
 0&=\nabla^2\phib,
 \\
\eqlab{phib_BC_tot}
  \phib & =\Veo w(\sss_0)+\frac{\ii}{\kappa}\frac{\omgD}{\omgeo}\pp_\perp \phib,\;
 \sss_0\in\Omega_\mr{eo}
 \\
 \eqlab{v2tot_gov_bc}
 \langle \vvv_2\rangle&=\vvvSlipAC+\vvvSlipEO,\; \text{ for } \rrr = \sss_0.
 \eal
 \esub
One could worry about using the electroosmotic slip velocity for discrete electrodes, where the edges of these equipotential surfaces introduce length scales that violate the assumptions necessary to derive \eqref{electro_eff_BC} and \eqref{EO_slip}. A two-dimensional Debye layer forms at the electrode edges, when the linearized system of equations \eqref{electrokinetic_gov_lin} is solved with boundary conditions \eqref{wall_pot} and \eqref{no_flux_lin} and a fully resolved boundary layer. This error only occurs at a relatively small part of the computational domain as long as $W_\mr{elec}\gg \lambdaD$. It was checked numerically, that the implementation of the effective boundary conditions only lead to a relative error of $0.8\,\%$ in the streaming pattern for a simple simulation with only two electrodes of widths $W_\mr{elec}=10\,\SImum$.

The 2D device simulation was performed for an increasing number of electrodes ranging from $N_\mr{elec}=6$ with $W_\mr{elec}=41.7\,\SImum$ to $N_\mr{elec}=25$ with $W_\mr{elec}=10.0\,\SImum$. For each value of $N_\mr{elec}$, the voltage was changed to minimize streaming at $E_\mr{ac}=100\,$Pa. This in turn required a decreasing amplitude for the applied voltage from $\Veo=222\,$mV at $N_\mr{elec}=6$ to $\Veo=162\,$mV at $N_\mr{elec}=25$. Unsurprisingly, it requires a higher voltage to generate streaming through the discrete electrode pattern compared to the idealized mode.

In \figref{Nelec_sweep} we plot the quantitative measures of the obtained streaming suppression versus the number $N_\mr{elec}$ of electrodes in the arrays. Notably, the initial increase from $6$ to $10$ electrodes yield the largest increase in the suppression measures, after which a gradual saturation sets in. This suggests that one can look for a reasonable trade off between having many electrodes and reaching a high suppression.

The simulated suppressed streaming is shown in \figref{Discrete_electrodes} for $N_\mr{elec} = 0,$ 6, 10, 15, 20 and 25. We notice, that the acoustic streaming at $N_\mr{elec}=0$ is almost identical to the idealized mode in \figref{Acoustic_mode}. For $N_\mr{elec}
=10$ the entire central part of the acoustophoretic channel is cleared for streaming amplitudes above $5\,\%$ of $v^\mr{Rayl}$. At $N_\mr{elec}=15$, a similar result is seen for the $2\,\%$ contour lines. Further increases in the electrode count largely just brings the contour lines closer to the boundaries.
It is likely undesirable to perform particle focusing close to the boundaries regardless, as this brings adhesion effects into the problem, so increasing the number of electrodes beyond a certain point appears redundant.
Right at the boundaries, high streaming velocities are still seen, but these decay on the length scale of the electrode widths. This is expected from the classical Rayleigh solutions, that finds an exponential decay in boundary driven streaming with a characteristic length scale identical to the parallel variations in the slip velocity.\cite{LordRayleigh1884}

\section{Discussion and conclusion}
\seclab{Discussion}
We have already addressed many of the technicalities associated with the method of combining acoustic and electroosmotic streaming. Here, before concluding, we will briefly address two critical points which were neglected in the preceding analysis. First, our analysis of electroosmotic streaming was based on the assumption of a vanishing intrinsic zeta potential. However, the presence of chemically generated charge on relevant interfaces like water/glass is inevitable. Typical values of the zeta potential for borosilicate glass is of the order of $\zeta\sim -100$~mV, \cite{Kirby2004} comparable in amplitude to the applied potentials. It is unclear whether or not a DC-offset of this intrinsic wall potential at the discrete electrode arrays would fully nullify its presence, as chemical charges would still form a significant charge cloud in the gaps between electrodes. One idea could be to make these gaps as small as possible.

The intrinsic zeta potential is typically modeled as a constant surface potential, and in the framework of the linearized theory, this would act as a zeroth order field, yielding non-zero gradients in $c_0$ and generating a corresponding initial steady electric equilibrium potential $\phi_\mathrm{0,fl}$. As noted in Ref.~\onlinecite{Olesen2006b}, the presence of these extra fields will lower the acquired slip-velocity. A supplemental numerical study in the linearized regime by the authors suggest, that the inclusion of a relatively high intrinsic zeta potential of $\zeta=-100\,$mV almost halves the slip velocity of the solution used in this paper with the shape remaining the same.

Secondly, we only worked with two-dimensional systems in our present study. While the electroosmotic problem could in principle be implemented invariant along the channel, we know from previous experimental studies,\cite{Augustsson2011} that even  long, straight microchannel with rectangular cross sections exhibits axial inhomogeneities in the acoustic fields, which renders a 3D analysis necessary for complete characterization of the system. Such a break in the 2D symmetry may yield areas of non-suppressed streaming, which could compromise the performance of the suggested chip design.

In conclusion, we have presented the theoretical framework for a method to effectively suppress acoustic streaming by superposing electroosmotic streaming. We have suggested a specific set of boundary conditions that achieves this suppression in a typical microchannel with a resonant standing half-wave. This idealized electroosmotic mode was then tested in an idealized hard-walled model of the microchannel, in a more realistic model including the transducer, the elastic solid and the microchannel, and for applied voltages in both the linear and nonlinear regime.

Furthermore, we have demonstrated that the electroosmotic streaming pattern derived from a linearized theory largely hold true for relevant amplitudes of applied voltages. Lastly, we also evaluated numerically the capability for  suppressing streaming of a specific acoustophoretic chip, where integrated discrete electrodes for generating electroosmosis were implemented. We hope that this theoretical work will inspire the acoustofluidic community to investigate experimentally the possibility of suppressing acoustic streaming in a controlled manner by electroosmosis.\\[-5mm]


\end{document}